\documentclass[aps,prd,tightenlines,preprintnumbers,showpacs,superscriptaddress,nofootinbib,eqsecnum,floatfix,amsmath,longbibliography,twocolumn]{revtex4-2}

\usepackage{bm}
\usepackage{latexsym}
\usepackage{graphicx}
\usepackage{multirow,soul}
\usepackage[dvipsnames]{xcolor}

\usepackage{hyperref}

\def\b{\beta}

\def\g{\gamma}

\def\m{\mu}
\def\n{\nu}

\def\cbo{{\,\raise-.15ex\Sc [\,}}                       

\def\svev#1{\left\langle #1\right\rangle}       

\def\ddt#1{{\buildrel {\hbox{\LARGE .\kern-2pt.}} \over {#1}}}

\long \def \blockcomment #1\endcomment{}

\definecolor{red}{rgb}{1., 0., 0.}
\definecolor{myred}{rgb}{0.8,0.0,0.0}
\definecolor{green}{rgb}{0.0,0.6,0.0}
\definecolor{darkblue}{rgb}{0.0,0.1,0.7}
\definecolor{brown}{rgb}{0.6,0.1,0.0}
\definecolor{gray}{rgb}{0.6,0.6,0.6}
\definecolor{darkgreen}{rgb}{0.0, 0.545098, 0.0}
\definecolor{verydarkgreen}{rgb}{0.0, 0.4, 0.0}
\definecolor{veryverydarkgreen}{rgb}{0.0, 0.3, 0.0}
\definecolor{purple}{rgb}{0.5,0.0,0.5}
\definecolor{applegreen}{rgb}{0.55, 0.71, 0.0}
\definecolor{babypink} {rgb}{0.64, 0.44, 0.44}
\definecolor{orange}{rgb}{0.9,0.4,0.0}

\def\ggf{g^2}
\def\svev#1{\left\langle #1\right\rangle}

\def\Eq#1{Eq.~(\ref{#1})}

\def\Fig#1{Fig.~\ref{#1}}

\def\betaS{\beta_\text{S}}
\def\betaW{\beta_\text{W}}

\def\sigmaS{\sigma_\text{S}}

\begin{document}

\title{ Infrared fixed point of the SU(3) gauge theory with $N_f = 10$ flavors }
\author{Anna Hasenfratz}
\affiliation{Department of Physics, University of Colorado, Boulder, CO 80309, USA}

\author{Ethan T. Neil}
\affiliation{Department of Physics, University of Colorado, Boulder, CO 80309, USA}

\author{Yigal Shamir}
\affiliation{Raymond and Beverly Sackler School of Physics and Astronomy,
Tel~Aviv University, 69978 Tel~Aviv, Israel}

\author{Benjamin Svetitsky}
\affiliation{Raymond and Beverly Sackler School of Physics and Astronomy,
Tel~Aviv University, 69978 Tel~Aviv, Israel}

\author{Oliver Witzel}
\affiliation{Center for Particle Physics Siegen, Theoretische Physik 1,
  Naturwissenschaftlich-Technische Fakult\"at, Universit\"at Siegen,
  57068 Siegen, Germany}


\begin{abstract}
We use lattice simulations and the continuous renormalization-group method, based on the gradient flow, to calculate the $\beta$ function and anomalous dimensions  of the SU(3) gauge theory with $N_f=10$ flavors of fermions in the fundamental representation.
We employ several improvements to extend the range of available renormalized couplings, including the addition of heavy Pauli-Villars bosons to reduce cutoff effects and the combination of a range of  gradient flow transformations.
While in the weak coupling regime our result is consistent with those of earlier studies,
our techniques allow us to study the system at much stronger couplings than previously possible.
We find that the renormalization group $\beta$ function develops a zero, corresponding to an infrared-stable fixed point, at gradient-flow coupling $\ggf=15.0(5)$.
We also determine the mass and tensor anomalous dimensions: At the fixed point we find $\gamma_m\simeq0.6$, suggesting  that this system might be deep inside the conformal window.
\end{abstract}

\preprint{SI-HEP-2023-14}

\maketitle

\newpage
\section{\label{sec:intro} Introduction}

The SU(3) gauge theory with ten Dirac fermions in the fundamental representation is the subject of continuing debate.
The question is whether its infrared physics is confining or conformal, as determined by the absence or presence of an infrared fixed point (IRFP).
The system  has been studied by several groups, both with domain wall \cite{Chiu:2016uui,Chiu:2017kza,Chiu:2018edw,Hasenfratz:2017qyr,Hasenfratz:2020ess} and staggered fermions \cite{Fodor:2018tdg,Fodor:2019ypi,Kuti:2022ldb}.
All these studies have used the finite-volume gradient flow (GF) scheme with a step-scaling renormalization-group transformation \cite{Luscher:2010iy,Luscher:2013cpa,Fodor:2012td}.
While the results are in reasonable agreement at weak gauge couplings, they differ at stronger couplings and reach differing conclusions.

\begin{figure}[htb]
\begin{center}
\includegraphics*[width=0.95\columnwidth]{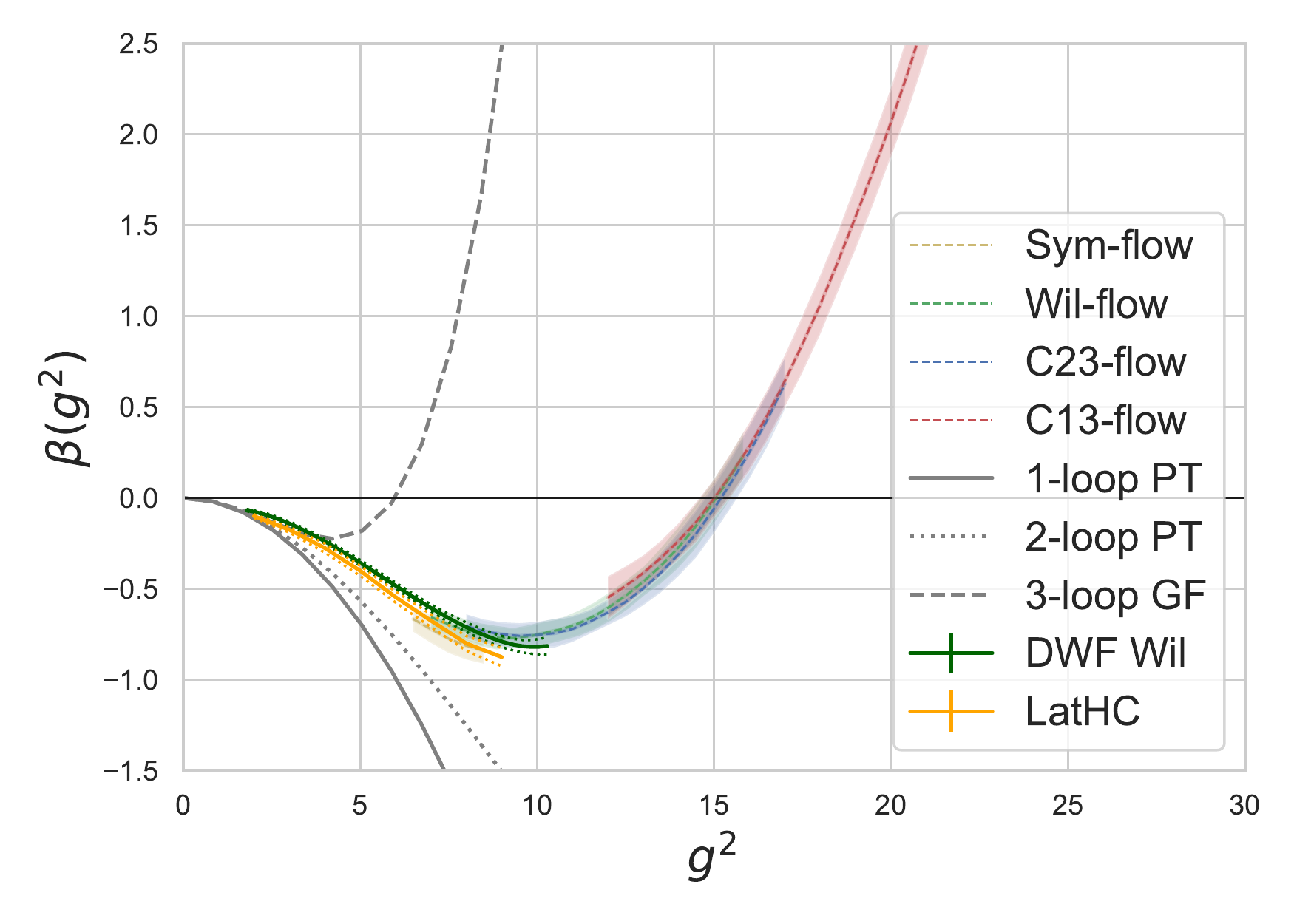}
\end{center}
\begin{quotation}
\caption{\label{betafn-final}
The $\beta$ function
obtained with four different gradient-flow transformations in overlapping regions.
The orange and dark green solid lines, with errors indicated by the dotted lines, are results from staggered and DWF simulations \cite{Fodor:2018tdg,Fodor:2019ypi,Kuti:2022ldb,Hasenfratz:2017qyr,Hasenfratz:2020ess,CBFinprogress}. Black solid, dotted and dashed curves correspond to the universal 1- and 2-loop and the gradient flow 3-loop perturbative results \cite{Harlander:2016vzb}.
}
\end{quotation}
\end{figure}
Using domain-wall fermions, Chiu \cite{Chiu:2016uui,Chiu:2017kza} first claimed an IRFP at $g^2\simeq 7$.
His later study \cite{Chiu:2018edw}, however, gave a more cautious assessment.
Hasenfratz, Rebbi, and Witzel \cite{Hasenfratz:2017qyr,Hasenfratz:2020ess} observed a step-scaling $\beta$ function that increases in absolute value up to $g^2\simeq 9$, where it appears to turn towards the abscissa and thus hints at an IRFP at some $g^2\gtrsim {11}$; these simulations were limited by a first-order phase transition blocking access to the $g^2 > 11$ regime.
Staggered-fermion calculations by the LatHC collaboration \cite{Fodor:2018tdg,Fodor:2019ypi,Kuti:2022ldb} studied this system in larger volumes, reaching couplings up to $g^2\simeq 10$.
In this range  their $\beta$ function increases steadily in magnitude and remains in $2\sigma$ agreement with the result reported in Refs.~\cite{Hasenfratz:2017qyr,Hasenfratz:2020ess}.
It does not, however, show any  sign of a developing IRFP\@.
No definitive conclusion on the infrared behavior of the $N_f=10$ model has been reached so far.

All the studies listed above were carried out in a range of renormalized coupling limited by large cutoff effects.
Recently we have proposed adding heavy Pauli-Villars (PV) bosons  to remove ultraviolet fluctuations caused by the many fermion fields \cite{Hasenfratz:2021zsl}.
The masses of the PV bosons are kept at the cutoff scale.
Thus they decouple in the continuum limit but they do generate a local effective gauge action with well-regularized short-distance properties.
We have applied PV improvement successfully in the SU(3) gauge theory with
$N_f=12$ \cite{Hasenfratz:2021zsl} and $N_f=8$ \cite{Hasenfratz:2022qan} staggered fermions in the fundamental representation, as well as in a multirepresentation SU(4) gauge theory with Wilson fermions \cite{Hasenfratz:2023sqa}. In all cases we found that the PV improved actions indeed reduced short-distance fluctuations and allowed investigations at stronger renormalized couplings.

In Ref.~\cite{Hasenfratz:2023sqa} we applied the continuous $\beta$ function (CBF) method \cite{Hasenfratz:2019hpg,Hasenfratz:2019puu} and uncovered an IRFP at strong coupling.
Going beyond the use of PV bosons,
we further extended the coupling range by combining the results of a number of lattice gradient flows   that possess a common continuum limit.
Here we apply the techniques used in \cite{Hasenfratz:2023sqa} to the SU(3) gauge theory with $N_f=10$ fundamental  flavors, simulated using Wilson fermions.
The presence of heavy PV bosons permits study of the system at much stronger renormalized couplings than previously possible.
When the CBF method is extended by varying the GF transformation, the accessible range of renormalized coupling is yet larger.
We summarize our main result---the $\beta$ function of the theory---in Fig.~\ref{betafn-final}.
By combining  overlapping results from different gradient flows, we  cover the range $6.5 \lesssim g^2 \lesssim 23$.
At weak coupling our prediction is consistent with that of LatHC~\cite{Fodor:2019ypi,Kuti:2022ldb} and overlaps with an CBF result \cite{CBFinprogress} obtained by reanalyzing the M{\"o}bius DWF data from Refs.~\cite{Hasenfratz:2017qyr,Hasenfratz:2020ess} (labeled ``DWF Wil'' in the figure).
 Our $\beta$ function turns around at $g^2\simeq 10.0$,  confirming the hint of an IRFP reported in \cite{Hasenfratz:2020ess}.
 At stronger coupling, the $\beta$ function rises steadily and crosses zero at an IRFP at $g_\text{FP}^2=15.0(5)$, implying the theory  is infrared conformal.

This paper is organized as follows.
In  Sec.~\ref{tech} we briefly describe the numerical simulations and define the gradient flow transformations.
In the following sections we show how we obtain our results and extrapolate them to the infinite-volume and continuum limits.
For the $\beta$ function see Sec.~\ref{betafn} and for anomalous dimensions of fermion bilinears see Sec.~\ref{sec:andim}.
With the exception of the volume extrapolation, the techniques used have been explained at length in our recent paper \cite{Hasenfratz:2023sqa} and hence our presentation here is brief.
We discuss our results further in Sec.~\ref{sec:conc}.

\section{\label{tech} Simulation and gradient flow}

Our data emerge from four-dimensional configurations of the Euclidean gauge theory, generated by hybrid Monte Carlo simulations of the lattice theory.
Our lattice action couples the gauge field to Wilson-clover fermions after smearing the links with normalized hypercubic (nHYP) smearing \cite{Hasenfratz:2001hp,Hasenfratz:2007rf}.
The clover coefficient is $c_{SW}=1$ \cite{Bernard:1999kc,Shamir:2010cq} and the smearing parameters $\alpha_i$ are the original set (0.75, 0.6, 0.3).
The plaquette gauge action is supplemented by a term for nHYP dislocation suppression (NDS) \cite{DeGrand:2014rwa}.
To further tame gauge field roughness, we add
30 Pauli-Villars (bosonic Dirac) fields---three for each fermion flavor---with bare mass $am_{PV}=1$~\cite{Hasenfratz:2021zsl}.
For each lattice coupling $\beta_0$ we set the hopping parameter $\kappa\simeq\kappa_c$ so that the fermion mass, calculated from the axial Ward identity, is bounded by $|am_f|<5\times10^{-4}$.
The ensembles are listed in Table~\ref{ensembles}.

\begin{table}[t]
\begin{ruledtabular}
\begin{tabular}{clcc}
  $\beta_0$ & $\kappa$ & $L/a=24$ & $L/a=28$   \\
\hline
 6.0  & 0.12742 & 135 & 100 \\
 6.2  & 0.12695 & 100 & 100 \\
 6.3  & 0.12677 & 150 & 100 \\
 6.5  & 0.1265  & 155 & 100 \\
 6.7  & 0.1263  & 135 & \phantom{0}90 \\
 7.0  & 0.12612 & 160 & 110 \\
\end{tabular}
\end{ruledtabular}
\caption{\label{ensembles} List of the ensembles, showing the lattice coupling constant $\beta_0$, the hopping parameter $\kappa$, and the number of configurations with volume $L^3\times(2L)$ for the two volumes.
}
\end{table}

We extract the $\beta$ function and anomalous dimensions
using a continuous RG transformation based on gradient flow
\cite{Carosso:2018bmz,Fodor:2017die,Hasenfratz:2019hpg,Hasenfratz:2019puu,Hasenfratz:2023sqa,Peterson:2021lvb,Hasenfratz:2023bok,Fodor:2017die,Kuti:2022ldb}.
The RG length scale is given by $\sqrt{t}$ where $t$ is the  GF time  \cite{Luscher:2010iy}.
In continuum language, the GF running coupling at scale $\sqrt t$ is defined as
\begin{equation}
\label{ggf}
\ggf = \frac{{\cal N}}{(1+\delta)} t^2 \svev{E(t)} \ ,
\end{equation}
with the energy density $E=\frac{1}{4}G_{\m\n}^a G_{\m\n}^a$, calculated from the flowed gauge field strength $G_{\m\n}^a$.
The constant ${\cal N} = 128\pi^2/(3(N_c^2-1))$ is a
normalization factor to match to the 1-loop $\overline{\text{MS}}$ result, while $\delta$ is a small finite-volume correction
\cite{Fodor:2012td}.
The average in \Eq{ggf} is over the ensemble of configurations generated at given lattice coupling $\beta_0$.
The $\beta$ function is simply
\begin{equation}
\label{betafn_def}
\b(\ggf) = -t\,\frac{\partial \ggf}{\partial t} \ .
\end{equation}
The anomalous dimensions (see below) are determined from the derivative of flowed correlation functions \cite{Carosso:2018bmz,Hasenfratz:2022wll,Hasenfratz:2023sqa}.

In the lattice theory, the GF transformation can be performed with different flows, each originating in a particular discretization of the gauge action.
We make use of four such lattice flows, corresponding to combinations of plaquette and rectangle terms with coefficients $c_p$ and $c_r$.
Imposing the perturbative normalization $c_p+8c_r=1$, we define the Symanzik flow ($c_p=5/3$), the Wilson flow ($c_p=1$), and flows called C23 ($c_p=2/3$) and C13 ($c_p=1/3$) \cite{Hasenfratz:2023sqa}.
The different flows correspond to different renormalized trajectories (RT) of the RG transformation.
Ideally, we would choose the RG transformation whose RT is closest to the simulation action.
We find that GF with  smaller $c_p$ values  are better for this purpose at stronger gauge coupling.

Besides this, the energy operator $E(t)$ can be defined by a variety of discretizations.
We calculate the W (Wilson), S (Symanzik), and C (clover) operators in order to gauge the approach of each to the continuum limit.
The S operator gives the smoothest approach, and so we will focus on results obtained using this operator.

\section{\label{betafn} The $\beta$ function}

\begin{figure}[t]
\begin{center}
\includegraphics*[width=0.95\columnwidth]{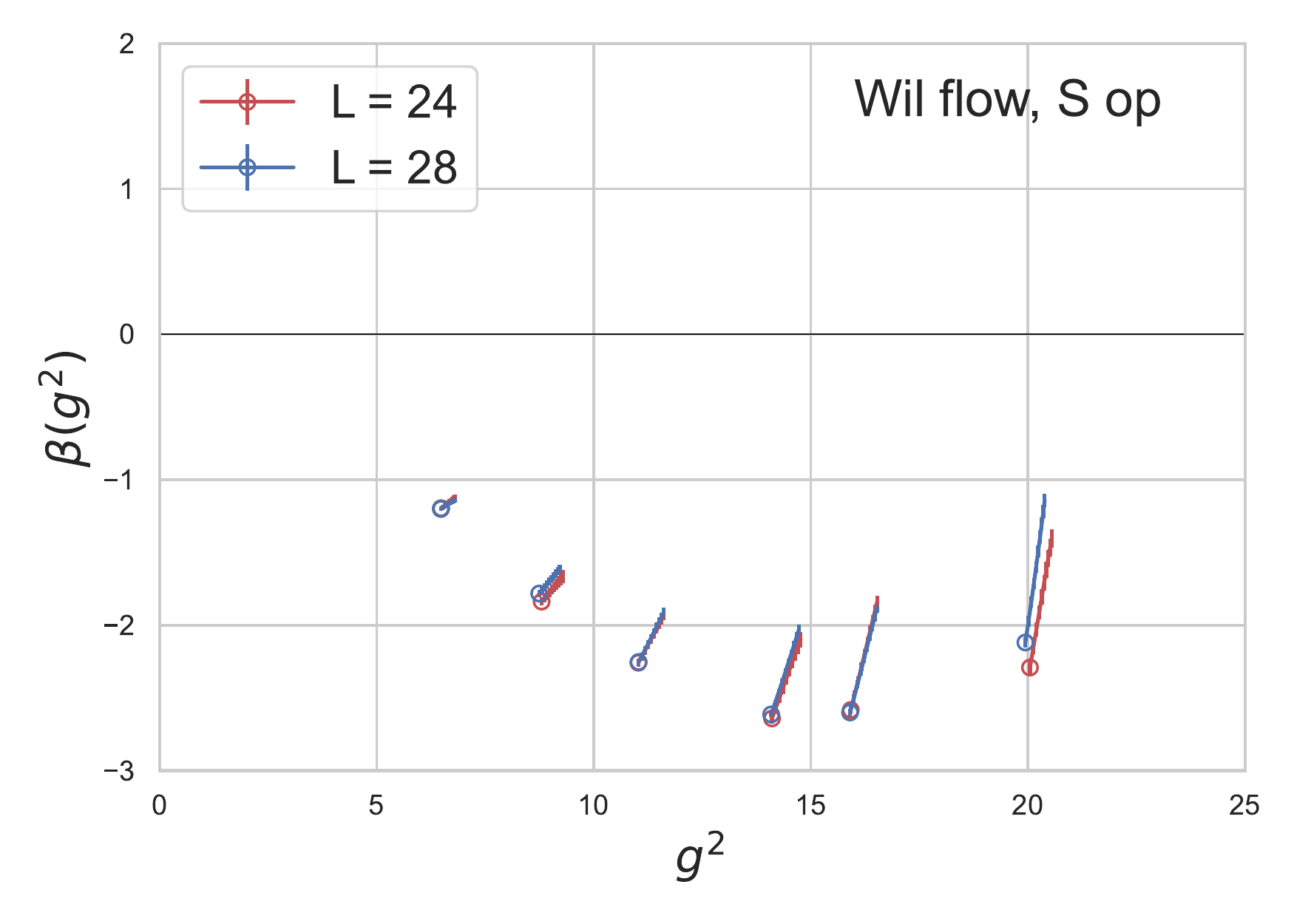}\\
\includegraphics*[width=0.95\columnwidth]{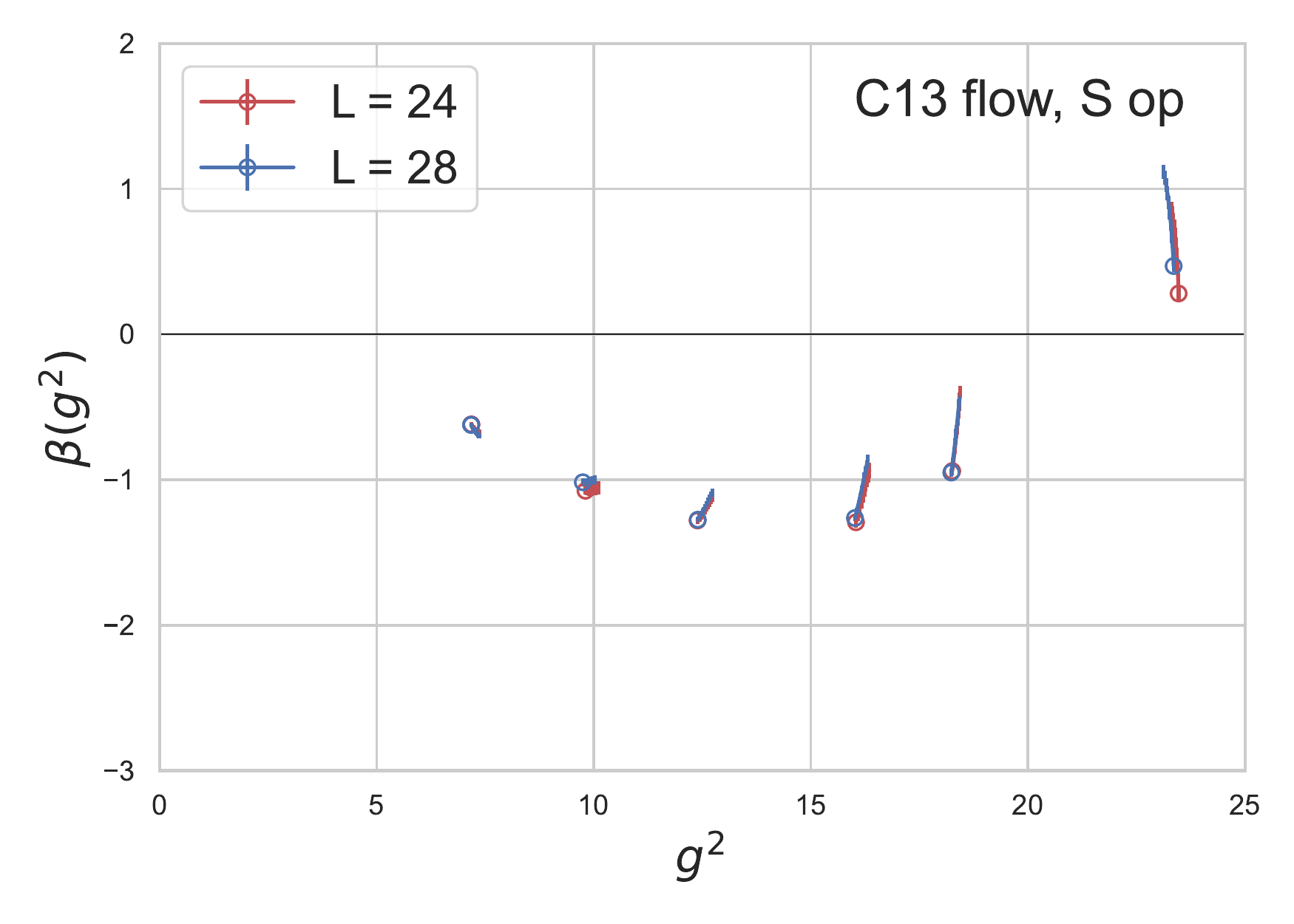}\vspace*{-4ex}
\end{center}
\caption{\label{fig:C13-raw}
Raw data for the flowed coupling $g^2$ and its derivative $\beta(g^2)$, calculated from the Symanzik operator.
The upper panel shows results for the Wilson flow; the lower, for C13 flow.
Each ensemble is represented by a group of data points:  From  right to left: $\beta_0=6.0$, 6.2, 6.3, 6.5, 6.7, 7.0.
In each group, the ensemble average at $t=2.8$ is plotted as a circle, from which rises the sequence of averages calculated at increasing $t$ for $2.8<t/a^2\leq3.8$, rising as $t$ grows. The lines at each point give a sense for the magnitude and direction of the continuum extrapolation $a^2 / t \rightarrow 0$.
The colors allow comparison of lattice size $L=24a$ (red) to $L=28a$ (blue).}
\end{figure}

Raw data for the flowed coupling $g^2$ and its derivative $\beta(g^2)$ from Wilson and C13 flows, are shown for all ensembles in Fig.~\ref{fig:C13-raw}.
One sees an upward flow with increasing $t$, ever faster for ensembles with larger coupling $g^2$.
The C13 flow has the shortest paths as $t$ grows, indicating that its RT is closest.
Moreover, while both flows show that the $\beta$ function approaches the axis for large $g^2$, the C13 flow actually shows a positive $\beta$ function at the strongest coupling, even before taking the continuum limit.
This is the first hint of an IRFP\@.
Finally, the figure allows comparison of lattice sizes $L^3\times (2L)$ for $L/a=24$ and $L/a=28$, whence it is seen that differences are very small but not zero.

We determine the continuous $\beta$ function via the following four steps, ultimately carrying out the limits $L\to\infty$ and $a\to0$ \cite{Hasenfratz:2019puu,Hasenfratz:2023bok}.

\subsection{ Infinite volume limit \label{sec:volume}}
\begin{figure}[t]
\begin{center}
\includegraphics*[width=0.95\columnwidth]{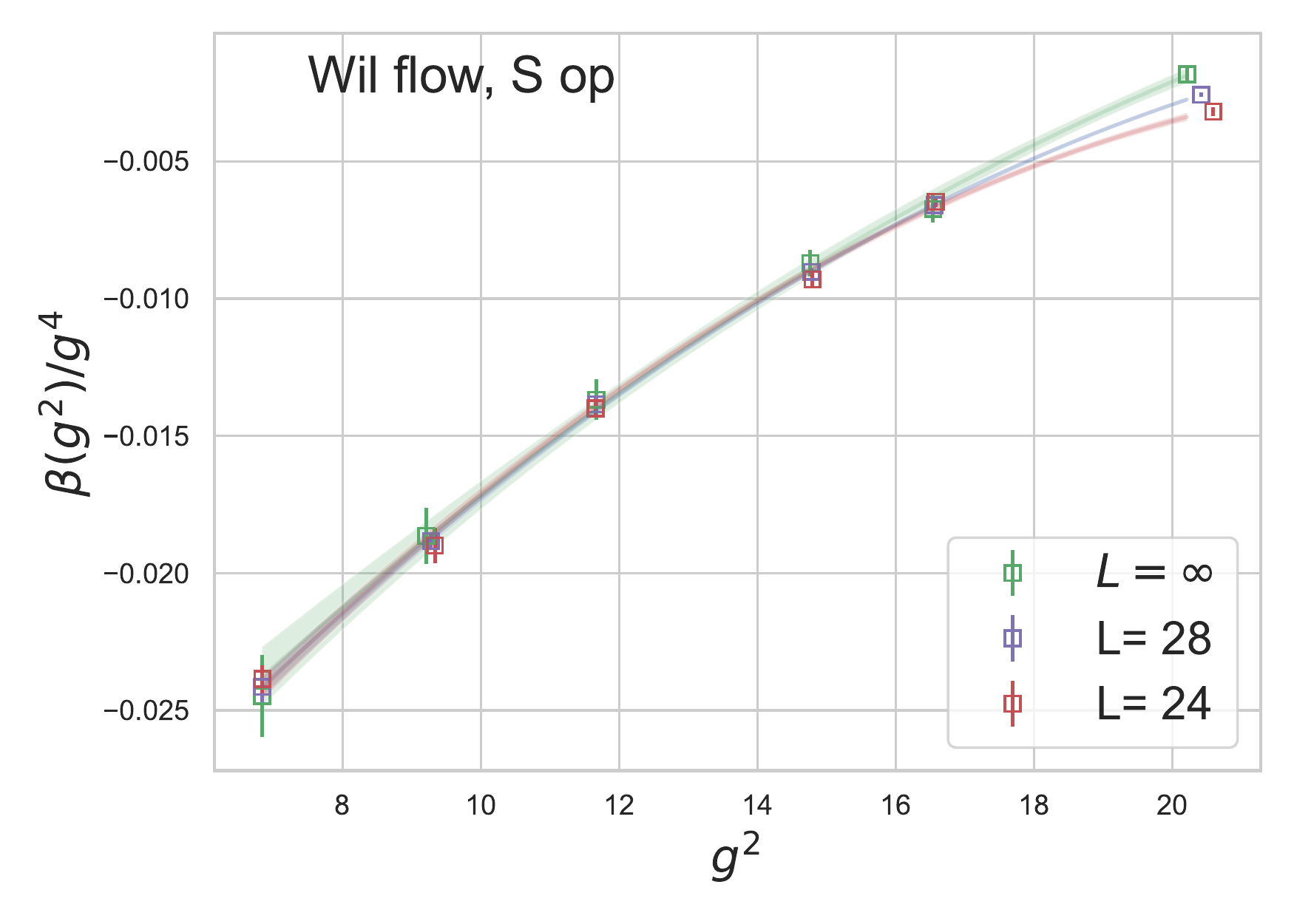}\vspace*{-4ex}
\end{center}
\begin{quotation}
\caption{\label{fig:interpolation-volume-dependence}
Comparing the interpolations of $L=24a$ (red) and $L=28a$  (blue), derived with Wilson flow, S operator at flow time $t/a^2=3.8$ (the largest flow time we consider). The green symbols show the extrapolation to infinite volume. The colored bands correspond to quadratic interpolations as explained in Sec.~\ref{interpolation}.
}
\end{quotation}
\vspace*{-3ex}
\end{figure}

We take the infinite volume limit, for both $\ggf$ and $\beta(\ggf)$, at each bare coupling $\beta_0$ and at selected flow times.
Simple scaling arguments imply that the leading volume dependence of the renormalized GF coupling is $\propto t^2/L^4$. Higher order corrections can hence be suppressed by choosing $t/L^2$ sufficiently small. In this study we consider two volumes, $L/a=24$ and 28, and restrict the flow time to $2.8\le t/a^2\le 3.8$. With this restriction the finite volume effects are well controlled in all ensembles.
In \Fig{fig:interpolation-volume-dependence} we plot $\beta(g^4)/g^4$, derived from Wilson flow at fixed flow time $t/a^2=3.8$, at all six bare coupling values (see Table~\ref{ensembles}) and both volumes.
The plot also shows the infinite volume extrapolation, assuming the leading order volume dependence is $1/L^4$.
The colored bands show quadratic interpolations of the data, which furnish values of $\beta(g^2)$ at intermediate values of $g^2$ (see below).
 In the next two steps we use only the infinite-volume extrapolations.

\subsection{Interpolation \label{interpolation}}
In order to take the the continuum limit  of  $\beta(\ggf; t/a^2)$, we need to determine pairs of ($\ggf$, $\beta(\ggf)$) at selected flow times for values of $\ggf$ between those that emerge from the ensembles.
As in the example of \Fig{fig:interpolation-volume-dependence}, we interpolate  $\beta(\ggf; t/a^2)$  with a quadratic form according to $\beta(g^2;t/a^2)/g^4=c_0+c_1 g^2 + c_2 g^4$ at a series of  flow times.
(See Ref.~\cite{Hasenfratz:2023sqa} for details.)
We do the same for the other flows.%
\footnote{The C13 flow shows strong cutoff effects at the weakest bare gauge coupling $\beta_0=7.0$, and hence we do  not include that value in the interpolation.
Similar phenomena were observed and discussed in Ref.~\cite{Hasenfratz:2023sqa}.}
As usual in interpolating data, we will not use the interpolating curves outside the range of the interpolation when taking the continuum limit.
The columns labelled $g^2_\text{min}$ and $g^2_\text{max1}$ in Table~\ref{tab:cuts} list the minimal and maximal values covered by the interpolating curves.

\subsection{ \label{contlim} Continuum limit}
\begin{figure}[tb]
\begin{center}
\includegraphics*[width=0.95\columnwidth]{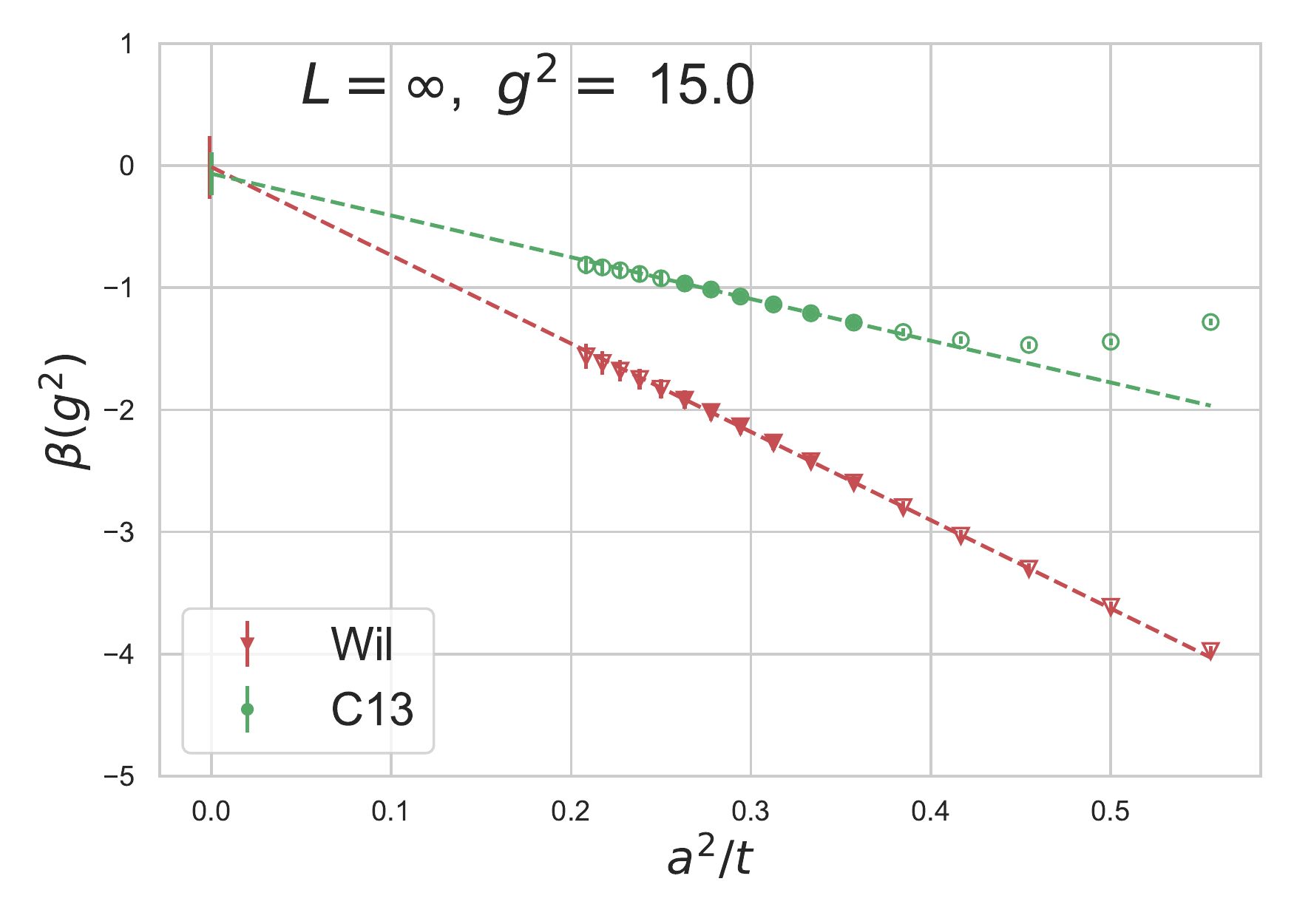}\\
\end{center}
\caption{\label{fig:cont-lim-Linf}
Continuum extrapolation $a^2/t\to0$ of the S operator for Wilson and C13 flows at $\ggf=15.0$ in the infinite-volume limit.
The solid symbols correspond to $2.8\leq t/a^2\leq3.8$, the range of flow time used in  the  extrapolations.
The curvature observed at small $t$ (large $a^2/t$) indicates that  the flow has not yet reached the renormalized trajectory.
}
\end{figure}

After determining interpolating curves as in \Fig{fig:interpolation-volume-dependence} for many values of $t/a^2$ in the interval $[2.8,3.8]$, we extrapolate  $\beta(\ggf; t/a^2)$ to $t/a^2 = \infty$ at fixed $\ggf$.
One such extrapolation is shown in \Fig{fig:cont-lim-Linf}.
Repeating this in a range of $g^2$ gives the $\beta$ function that is plotted in Fig.~\ref{betafn-final}.
We show curves and error bands for the results of Symanzik, Wilson, C23, and C13 flows.
These must agree in the continuum limit.
Figure~\ref{fig:cont-lim-Linf}, in particular, shows the excellent agreement of Wilson and C13 flows at $g^2=15.0$, and the fact that $\beta(g^2)$ is consistent with zero there.

It can be seen in \Fig{betafn-final} that each flow is plotted in a restricted range of $g^2$.
This stems from a requirement of internal consistency, as follows.

\subsection{ \label{consist} Consistency tests}
At any given physical coupling $\ggf$, the $\beta$ functions based on different discretizations of the flowed energy density---the S, W, and~C operators---must agree in the continuum limit.
For each flow, we use this requirement on the operators
to impose cuts on the range of $g^2$ where that flow can be trusted.

\begin{figure}[t]
\begin{center}
\includegraphics*[width=0.95\columnwidth]{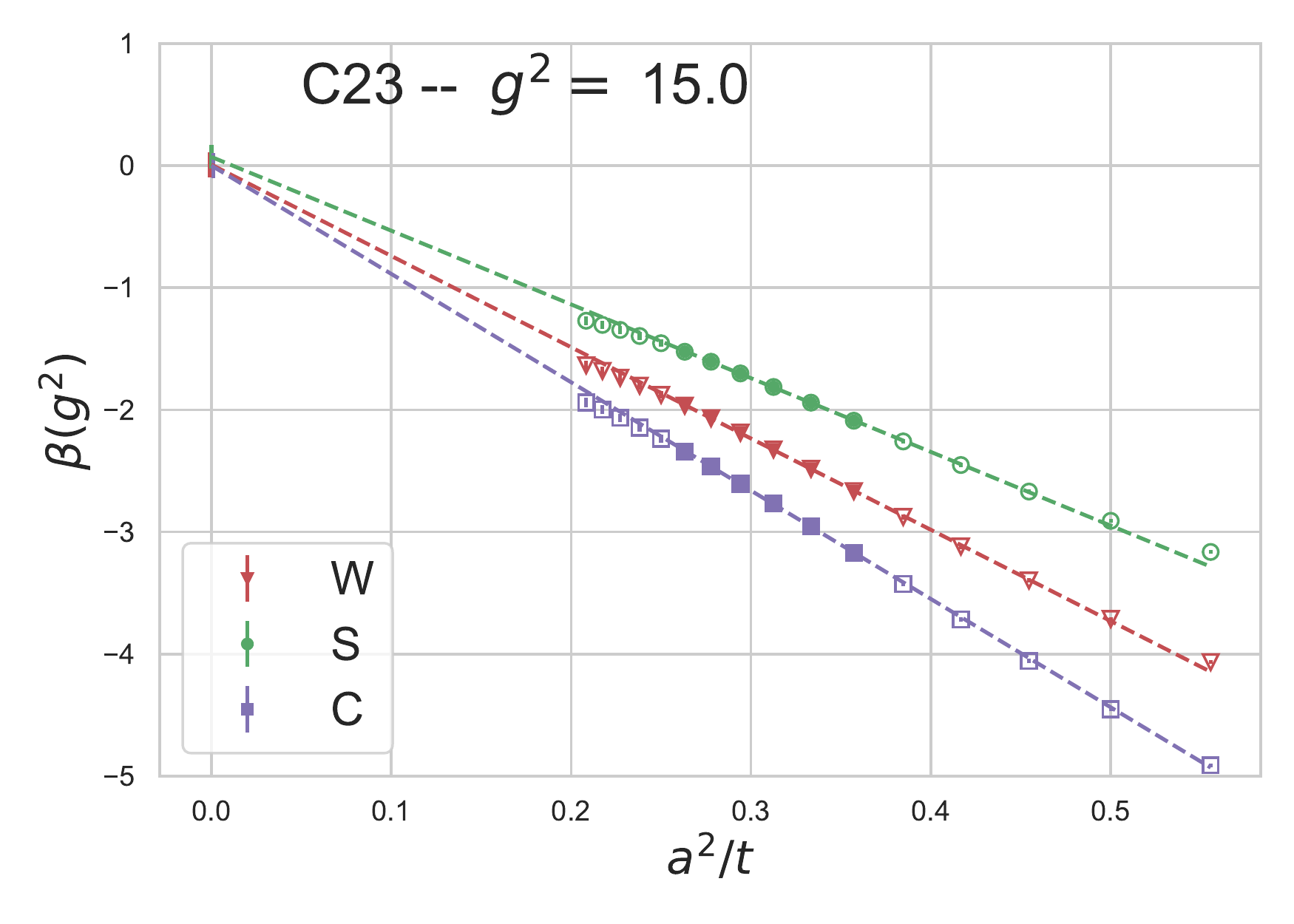}\vspace*{-4ex}
\end{center}
\caption{\label{fig:cont-lim}
Continuum limits $a^2/t\to0$ of data from the C23 flow on the single volume $24^3\times48$ (cf.~\Fig{fig:cont-lim-Linf}).
We show the extrapolation of data from S, W, and~C operators.
We use the S and W extrapolations to test for consistency.
The curvature observed in the data at large flow times (small $a^2/t$) indicates finite volume effects, absent from the $L\to\infty$ limit shown in \Fig{fig:cont-lim-Linf}.}
\end{figure}

We return to the continuum extrapolation of \Fig{fig:cont-lim-Linf},
but we step back to consider only a single volume (cf.~Fig.~\ref{fig:interpolation-volume-dependence}) where  we compare the results stemming from the three operators, S, W, and~C\@.
We show an example of these continuum extrapolations in \Fig{fig:cont-lim}.
For all flows and at every $\ggf$ we observe
that the S operator's extrapolation has the smallest slope, meaning the smallest cutoff effects, with the W operator coming next.
The C operator is furthest from the continuum limit, having the largest slopes.
This naturally leads to the choice of the S operator
for our main result, but we require consistency between the S and W operators.
We label the extrapolated values of the S and~W data at each value of $g^2$ as $\betaS$ and $\betaW$.

\begin{figure}[h]
\begin{center}
\includegraphics*[width=0.95\columnwidth]{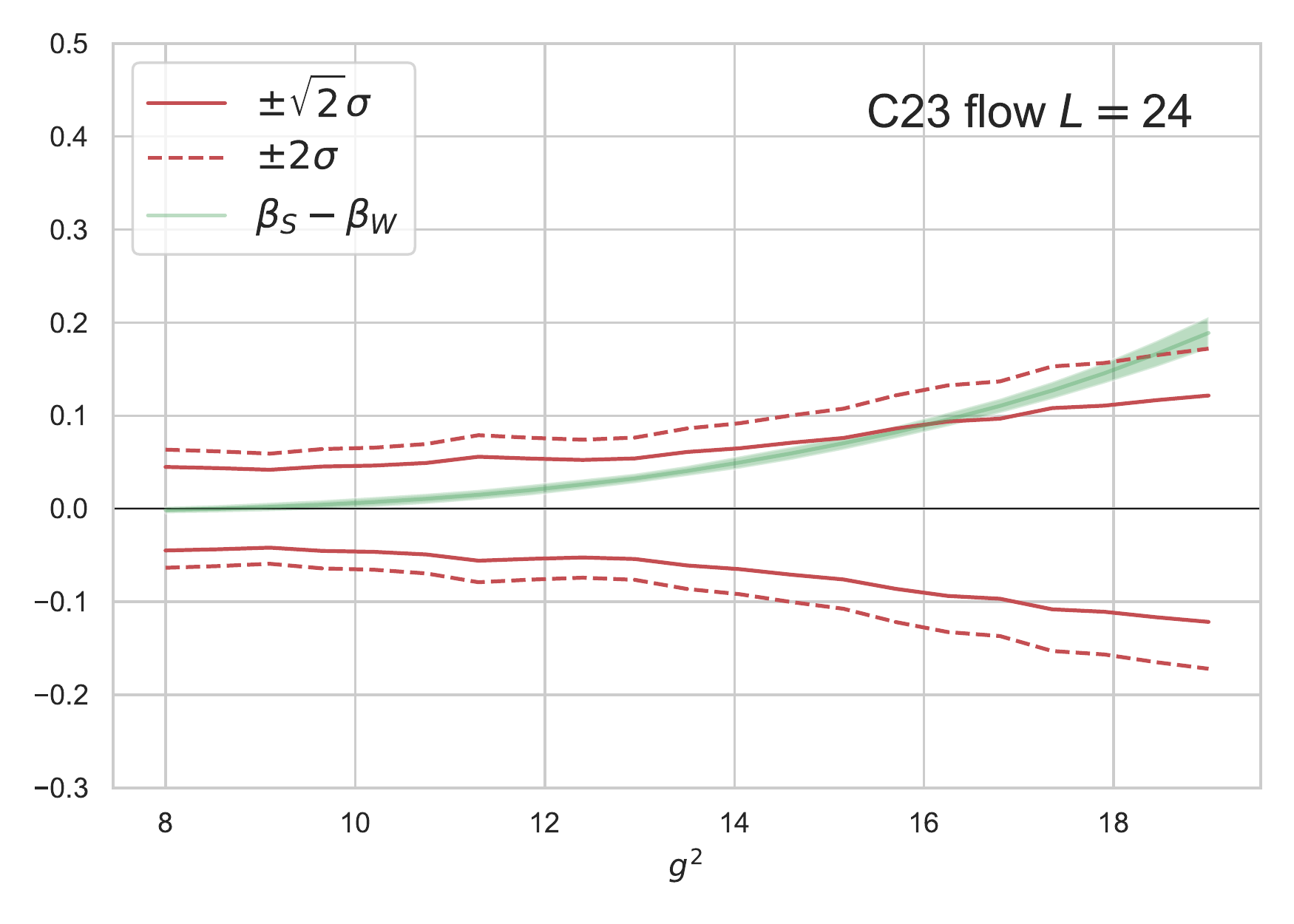}\vspace*{-4ex}
\end{center}
\caption{\label{fig:consistency} Comparing the difference $\beta_S-\beta_W$ (green)
with the standard deviation $\sigma$ of the S operator on volume $24^3\times48$
for  the C23 flow.
We plot $\pm\sqrt2\sigma$ (solid red curves) and $\pm2\sigma$ (dashed red curves).
Together with a similar plot for volume $28^3\times56$, these determine
the bounds of validity $g^2_\text{max3}$ and $g^2_\text{max2}$ in Table~\ref{tab:cuts}.
}
\end{figure}

\begin{figure}[htb]
\begin{center}
\includegraphics*[width=0.95\columnwidth]{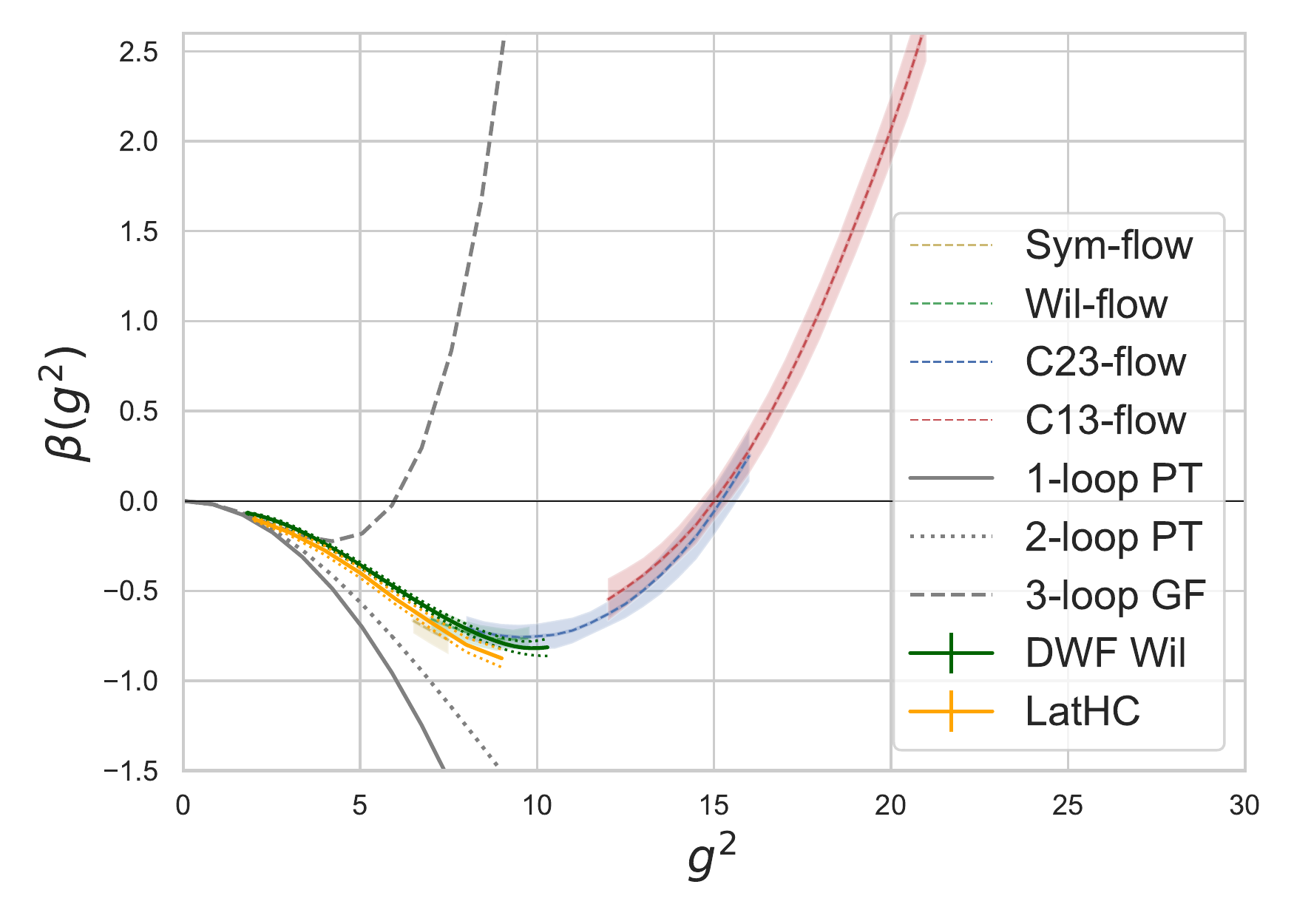}\vspace*{-4ex}
\end{center}
\caption{\label{betafn-strict}
Same as \Fig{betafn-final}, but with stricter bounds on the domain of validity of each flow.
See Table~\ref{tab:cuts}.
}
\end{figure}

We base the criterion for consistency on plots like the one shown in \Fig{fig:consistency}, which is obtained from a bootstrap analysis and includes correlations among the different operators.
The green band is the $\pm1\sigma$ error band of the difference $\betaS-\betaW$.
The solid red curves represent $\pm\sqrt2\sigmaS$, where $\sigmaS$ is the error
in $\betaS$, while the dashed red curves are $\pm2\sigmaS$.
The green band is much narrower than the span between the red curves; this reflects the strong correlations between the operators.
For a  consistency test  we require   $|\betaS-\betaW|\lesssim  2 \sigma_S$ on $L=24$ and~28 volumes separately and simultaneously.
This requirement restricts the $g^2$ values where a given flow can be trusted.
Table~\ref{tab:cuts} lists the corresponding values for each flow.
The demand that $|\betaS-\betaW|\lesssim 2\sigma_S$ gives us the bounds for each flow that are listed in columns 2 and~4 in the table and shown in \Fig{betafn-final}.
A slightly stricter constraint is obtained by requiring $|\betaS-\betaW| \lesssim \sqrt{2} \sigma_S$.
This lowers some of the upper bounds, as shown in the table (see column 5) and  in \Fig{betafn-strict}.
In this case the Wilson flow no longer reaches the fixed point, but there is little effect on the C23 and C13 flows, which do.

\begin{table}[t]
\begin{ruledtabular}
\begin{tabular}{crrrr}
Flow     & $g^2_\text{min}$ & $g^2_\text{max1}$ &  $g^2_\text{max2}$ & $g^2_\text{max3}$
 \\ \hline
 Sym & 6.5 & 18.0 &  8.5 &  7.5 \\
 Wil & 7.0 & 20.0 & 16.0 & 10.0 \\
 C23 & 8.0 & 21.0 & 17.0 & 16.0 \\
 C13 &12.0 & 23.0 & 23.0 & 21.0 \\
\end{tabular}
\end{ruledtabular}
\caption{\label{tab:cuts} Ranges of $\ggf$ in which each flow is included in the final result
for $\beta(\ggf)$.  $g^2_\text{min}$ and $g^2_\text{max1}$ result
from  the interpolations,
while $g^2_\text{max2}$ and $g^2_\text{max3}$ come from further demanding
consistency between the continuum extrapolations $\betaS$ and $\betaW$
(see \Fig{fig:consistency}).
We quote all numbers with a resolution of 0.5.
$g^2_\text{min}$ and $g^2_\text{max2}$ give the ranges reflected in \Fig{betafn-final}.
$g^2_\text{max3}$, rather than $g^2_\text{max2}$, gives the stricter bounds shown in \Fig{betafn-strict}.
}
\end{table}

\section{\label{sec:andim} Anomalous dimensions}

The calculation of anomalous dimensions follows that of Ref.~\cite{Hasenfratz:2023sqa} closely, with the addition of an extrapolation to infinite volume as described in Sec.~\ref{sec:volume} above.
We calculated the two-point function of each flowed mesonic density $X'$ with its unflowed source $X$,
\begin{equation}
\label{meson}
\svev{X(0)\, X'(t)} \sim t^{-(d+\eta+\g)/2} \ .
\end{equation}
Here $\gamma$ is the desired anomalous dimension of the operator and $\eta/2$ is the anomalous dimension of the the elementary fermion field.
To eliminate $\eta$,
we divide $\svev{X(0)\, X'(t)}$ by the two-point function of the  conserved vector current.
Defining the ratio
\begin{equation}
\label{Rratio}
  R(t) = \frac{\svev{X(0)\, X'(t)}}{\svev{V(0)\, V'(t)}}\ ,
\end{equation}
we have
\begin{equation}
\label{R}
  R(t)  \sim t^{-\g/2} \ ,
\end{equation}
and hence $\g$ can be extracted from the logarithmic derivative,
\begin{equation}
\label{gammaX}
\g = -2\frac{t}{R}\,\frac{\partial R}{\partial t} \ .
\end{equation}
We require $\sqrt{8t}\ll x_4$, where $x_4$ is the separation of $X$ and~$X'$ in Euclidean time.
This means that $x_4$ is kept large compared to the smearing of the operators by the flow.
The extrapolation from $L/a=24$, 28 to $L=\infty$,  the interpolation in $g^2$ at fixed $t$, and the continuum extrapolation $t/a^2\to\infty$ are as described above and in Ref.~\cite{Hasenfratz:2023sqa}.

\begin{figure}[tb]
\begin{center}
\includegraphics*[width=0.95\columnwidth]{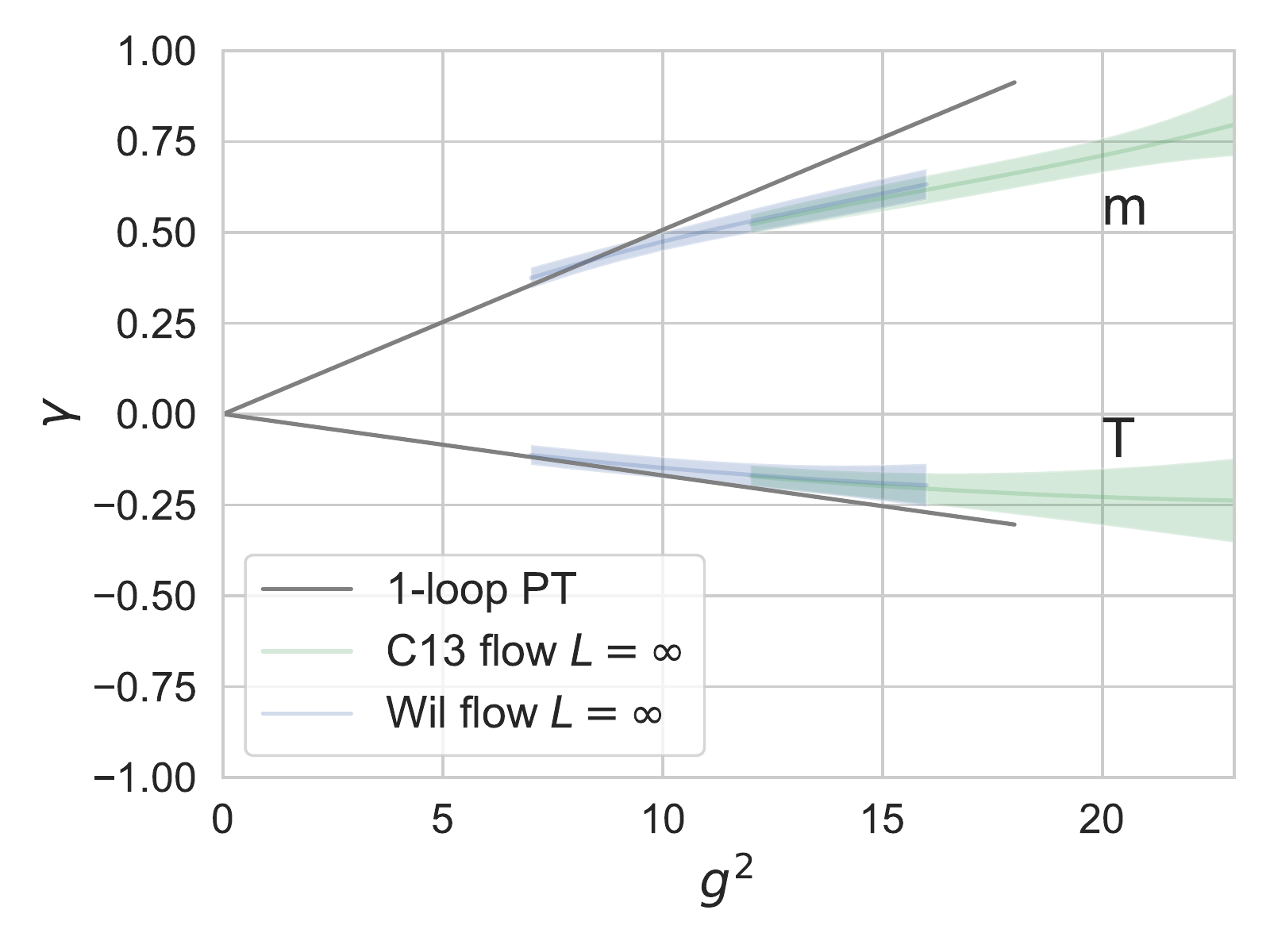}\vspace*{-4ex}
\end{center}
\caption{\label{gamma-final}
The anomalous dimension  of the mass (scalar) operator, $\gamma_m$, and that of the tensor operator, $\gamma_T$, obtained with Wilson and C13 flows, extrapolated to the continuum limit and to infinite volume.
}
\end{figure}

Final results for the mass anomalous dimension and for that of the tensor density are shown in \Fig{gamma-final}.
In the weak-coupling region, the anomalous dimensions
agree with one-loop perturbation theory,
\begin{equation}
\label{gamma1loop}
\g_m = \frac{6 \ggf C_2}{16 \pi^2} \ ,\quad \g_T = -\frac13\gamma_m\ ,
\end{equation}
where $C_2=4/3$ is the quadratic Casimir operator of the fermion representation.
At larger couplings, the magnitudes of both $\gamma_m$ and $\gamma_T$
drop below the respective one-loop results.
At the IR fixed point $\ggf\simeq 15$ we find $\g_m\simeq 0.6$
and $\g_T\simeq -0.2$.

The LSD collaboration~\cite{LatticeStrongDynamics:2020uwo,Witzel:2020hyr} calculated $\g_m$  in the 4+6 mass-split system using hyperscaling relations and reported $\g_m= 0.47(5)$.
That work  performed simulations near $\ggf\simeq 10$.
Accounting for the dependence of $\gamma_m$ on $g^2$ (cf.~\Fig{gamma-final}), the two values are consistent.

\section{\label{sec:conc} Conclusions}

We have presented a calculation of the $\beta$ function of the SU(3) gauge theory with $N_f=10$.
Our result is represented by a consistent set of  curves that derive from gradient-flow transformations with overlapping regions of validity, giving overall a smooth graph of the $\beta$ function.
Our work makes contact with existing results at weak coupling and reaches much larger couplings than previously attainable.

We obtain strong evidence for an infrared-stable fixed point at $g^2\simeq 15$, whose location is consistently identified by three different flows.
This places the theory inside the conformal window for SU(3) gauge theories with fermions in the fundamental representation.
We remind the reader that the Banks-Zaks fixed point%
 \cite{Banks:1981nn},  based on the two-loop $\beta$ function, places the sill of the window at $N_f=8.05$.

We have similarly calculated the anomalous dimensions $\gamma_m$ and $\gamma_T$.
As shown above, the mass anomalous dimension has the value $\gamma_m\simeq0.6$ at the fixed point, in good agreement with the value of 0.615
obtained by Ryttov and Shrock using their scheme-independent series expansion
to fourth order \cite{Ryttov:2016asb,Ryttov:2017kmx}.
Likewise, there is good agreement for the tensor anomalous dimension,
for which they obtained the value of $-0.149$ at third order \cite{Ryttov:2016hal}.

It is generally expected \cite{Kim:2020yvr} that $\g_m\to 1$ at the sill of the conformal window.
Our result suggests that the $N_f=10$ theory is well above the sill.

\begin{acknowledgments}
We thank R.~Shrock for correspondence about Refs.~\cite{Ryttov:2016asb, Ryttov:2016hal,Ryttov:2017kmx}.
Computations for this work were carried out on facilities
of the USQCD Collaboration, which are funded by the Office of Science
of the U.S.~Department of Energy.
A.H. and E.N.\ acknowledge support by DOE grant DE-SC0010005.
The work of B.S.\ and Y.S.\ was supported by the Israel Science Foundation
under grant No.~1429/21.
Our simulation code is a derivative of the MILC code \cite{MILC}.

\end{acknowledgments}

\bibliography{Nf10letter}         

\end{document}